\documentclass[12pt,tightenlines,eqsecnum,floats,showpacs,nofootinbib,amsmath,amssymb,aps,prd,superscriptaddress]{revtex4-1}

\usepackage{graphicx}
\usepackage{amsmath,amssymb}
\usepackage{hyperref}

\begin{document}
	
\title{Modeling effective FRW cosmologies with perfect fluids from states of the hybrid quantum Gowdy model}

\author{Beatriz Elizaga Navascu\'es}
\email{belizaga@estumail.ucm.es}
\affiliation{Universidad Complutense de Madrid, 28040 Madrid, Spain}
\affiliation{Instituto de Estructura de la Materia, IEM-CSIC, Serrano 121, 28006 Madrid, Spain}
\author{Mercedes Mart\'in-Benito}
\email{mmartin@hef.ru.nl}
\affiliation{Radboud University Nijmegen, Institute for Mathematics, Astrophysics and Particle Physics, Heyendaalseweg 135, NL-6525 AJ Nijmegen, The Netherlands}
\author{Guillermo A. Mena Marug\'an} \email{mena@iem.cfmac.csic.es}
\affiliation{Instituto de Estructura de la Materia, IEM-CSIC, Serrano 121, 28006 Madrid, Spain}
	
\begin{abstract}
We employ recently developed approximation methods in the hybrid quantization of the Gowdy $T^3$ model with linear polarization and a massless scalar field to obtain physically interesting solutions of this inhomogeneous cosmology. More specifically, we propose some particular approximate solutions of the quantum Gowdy model constructed in such a way that, for the Hamiltonian constraint, they effectively behave as those corresponding to a flat homogeneous and isotropic universe filled with a perfect fluid, even though these quantum states are far from being homogeneous and isotropic. We analyze how one can get different perfect fluid effective behaviors, including the cases of dust, radiation, and cosmological constant.
\end{abstract}
	
	
\maketitle
	
\section{Introduction}
\label{sec:Intro}

One of the most challenging tasks of theoretical physics is to provide a complete quantum theory for gravity and extract predictions from it. Without a quantum gravity theory we lack an understanding of the microscopic structure of the space-time, as well as a description of the very early stages of the Universe.

In recent years, cosmology has undergone remarkable progress, in particular thanks to the  precise measurements of the fluctuations of the primordial radiation, as imprinted in the cosmic microwave background (CMB) \cite{cmb}. Furthermore, new recent experiments have measured the influence that B-modes, which could have their origin on primordial gravitational waves, have on the polarization of the CMB  \cite{Bmodes}. Primordial gravitational waves may carry information about the quantum fluctuations of the early Universe geometry, and therefore their study opens a window to investigate quantum effects of gravity in cosmology. Thus, we may have in reach the possibility of testing some of our quantum gravity proposals contrasting their predictions with future cosmological data. 

Loop quantum gravity \cite{lqg} is one of the most solid approaches to attain the quantization of gravity. Its consequences in cosmology have been studied in the last decade, in particular within the line of research  known as loop quantum cosmology (LQC) \cite{lqc}. Initially, LQC focused on the study of homogeneous models. The most outstanding result is the resolution of  the cosmological singularity by means of a so-called quantum bounce (see, e.g., \cite{lqc} and references therein, especially \cite{bounce}). Even so, the extraction of realistic predictions from LQC calls for the study of inhomogeneous models. This fact motivated the analysis  in the framework of LQC of the simplest inhomogeneous cosmology, namely, the Gowdy model with the spatial topology of a three-torus, $T^3$. This model describes space-times with two axial Killing vectors \cite{gowdy} which contain gravitational radiation propagating in only one spatial direction. In the case of radiation with linear polarization, corresponding to just one of the two physical degrees of freedom of the gravitational waves, a hybrid approach was proposed for the quantization of the model in the context of LQC \cite{hybrid,hybrid3}. This  hybrid approach is based on a splitting of the phase space of the system in homogeneous and inhomogeneous sectors, and it implements a loop quantization for the homogeneous sector, whereas a more standard Fock quantization is adopted for the inhomogeneities. The approach therefore assumes that the most relevant effects of the quantization of the space-time geometry are those that affect the homogeneous degrees of freedom (at least in the regimes of interest in cosmology). 

Nowadays, with the aim of extracting physical predictions, the hybrid approach is being employed in the analysis of more realistic cosmological scenarios (free of Killing vector reductions), namely Friedmann-Robertson-Walker (FRW) models with cosmological perturbations \cite{inf-hybrid}, both in the geometry and in the matter content (see also \cite{inf-ash} for an alternate approach with a similar quantum treatment of the inhomogeneities, and e.g. \cite{effe, effe2} for an effective approach based on requirements about the closure of the constraint algebra). The quantum dynamics of these systems remain unsolved, owing mainly to the complexity of the Hamiltonian constraint, and only results employing  effective descriptions have been extracted. With the eye put on resolving the truly quantum dynamics of such systems, we keep investigating Gowdy models. In comparison with other more realistic scenarios, Gowdy cosmologies serve as a technically simpler arena where one can develop techniques, while still retaining the fundamental field-like features of inhomogeneous cosmologies. 

With this philosophy in mind, approximation methods were developed in \cite{hybrid-approx} for the  hybrid quantization of the Gowdy  model with linear polarization and $T^3$-topology in the case of local rotational symmetry (LRS), so that a single variable is enough to account for the anisotropies, and including as matter content a massless scalar field with the same symmetries as the geometry \cite{hybrid-matter}. This analysis made possible to obtain certain approximate solutions to the complicated Hamiltonian constraint of the model. Actually, this constraint is formed by four terms: the Hamiltonian constraint of a flat FRW model with three-torus spatial topology and coupled to a homogeneous massless scalar, an anisotropy term proportional to the momentum of the anisotropy variable (these two first terms in turn form the Hamiltonian constraint of a Bianchi I model \cite{Bianchi,awe1,MMP} with $T^3$-topology and LRS in the presence of a homogeneous massless scalar field), and two terms which couple the homogeneous sector with the inhomogeneities. One of these terms depends on the inhomogeneities via their free energy, and the other involves interaction among the different inhomogeneity modes while preserving the total field momentum. The inhomogeneities come both from the gravitational waves and from the massless scalar field. In this way we can regard the considered Gowdy model as a flat isotropic model on top of which we include anisotropies, as well as gravitational waves and matter inhomogeneities propagating in one direction. 

The discussion in \cite{hybrid-approx} provides states for which both the anisotropy term and the term involving the self-interaction of the inhomogeneities can be disregarded. These states are quite special and present a specific profile as far as the anisotropy variable is concerned. This profile is a Gaussian peaked on a large constant value of the anisotropy variable and on a vanishing value of its momentum. These states are approximate solutions of the Gowdy model whose peaks are far from describing homogeneous and isotropic space-times. Nevertheless, the quantum global effect of the anisotropies and inhomogeneities on the geometry is such that the states happen to correspond in turn to solutions of a flat FRW model with a massless scalar field. Indeed, on these states the constraint can be approximated by a constraint formed only by the FRW term and by the term involving the free energy of the inhomogeneities. This free energy term becomes a constant in such a simplified system. Moreover, this constant quantity can then be absorbed into the momentum of the homogeneous massless scalar field of the FRW model (which is also a constant of motion) by redefining it. 

The aim of the present paper is to extend this previous analysis to solutions for which the term containing the free energy of the inhomogeneities is no longer constant but evolves displaying a particular volume-dependent behavior. On the one hand, this extension shows that the family of states of the Gowdy model that can be regarded as approximate solutions of an FRW model is not as restrictive as one might infer from the discussion of \cite{hybrid-approx}. On the other hand, in doing this generalization, we want to construct states for which the mentioned volume-dependent term mimics the behavior of some perfect fluids, such as radiation, dust, cosmological constant, or other isotropic scalar field contents. We achieve this by still considering profiles for the anisotropy variable that are given by a Gaussian, but this Gaussian is now peaked on a volume-dependent value of the anisotropy variable, while retaining the property that this peak corresponds to a vanishing value of its momentum. The Hamiltonian constraint that each of these states verifies, which is an approximation to the exact Hamiltonian constraint, depends on the particular relation between the variables that appear as parameters in the definition of the peak of the Gaussian profile. We will analyze which kinds of dependence of the peak with the volume are acceptable and feasible, and how they can mimic different types of effective matter content. 

From the phenomenological point of view this is a pedagogically interesting analysis, as it materializes an example where quantum states that are genuinely anisotropic lead to isotropic descriptions, at least with respect to certain physical properties. This suggests the possibility that a similar study in the context of more realistic scenarios might shed light on the origin of the cosmological constant from an effective behavior of inhomogeneities for certain quantum states, providing a quantum mechanism to confront the long-standing cosmological constant problem \cite{cosmoconst}.

The structure of the paper is as follows. In Sec. \ref{sec:Gowdy} we summarize the hybrid quantization of the Gowdy $T^3$ model with linearly polarized gravitational waves, LRS, and a massless scalar field with the same symmetries as the geometry. This model was analyzed in  \cite{hybrid-matter}. Then, we review the approximate solutions constructed in \cite{hybrid-approx} which, as we have mentioned, lead effectively to isotropic universes with a massless scalar field as far as the Hamiltonian constraint is concerned. We also revisit the discussion about the regimes and approximations that justified these solutions. In Sec \ref{sec:mimic} we extend that analysis to volume-dependent Gaussian profiles with the aim of introducing approximate dynamical evolutions similar to those corresponding to other types of isotropic matter contents. We study under which conditions the approximations introduced in \cite{hybrid-approx} are still valid. We construct solutions verifying these approximations and analyze the kind of perfect fluid behavior that is allowed. Finally, in Sec. \ref{sec:conclu} we conclude with additional comments on the mathematical nature of our approximations and meaning of our approximate solutions, summarize our results, and discuss possible implications of them.

\section{Hybrid quantization of the Gowdy model and approximate solutions}
\label{sec:Gowdy}

\subsection{Quantization of the model and constraint operators}
\label{iia}

Let us review the hybrid quantization of the linearly polarized  Gowdy model with three-torus spatial topology, with LRS, and minimally coupled to a free massless scalar field $\Phi$ with the same symmetries as the metric \cite{hybrid-matter}.

Let $\{\theta, \sigma, \delta\}$ denote three orthogonal spatial coordinates, each of them defined on the circle. These coordinates are adapted to the symmetries so that matter and gravitational fields have spatial dependence only in e.g. $\theta$. Therefore, the fields can be expanded in Fourier modes in this coordinate. After fixing the gauge partially, the resulting reduced phase space consists of a homogeneous sector, which reproduces the phase space of a Bianchi I model with LRS and a minimally coupled homogeneous massless scalar field $\phi$ (the zero mode of the matter field $\Phi$ in the introduced Fourier expansion), and of an inhomogeneous sector, describing the non-zero modes of the linearly polarized gravitational waves and of the matter field, together with their canonically conjugate momenta. Two global constraints remain in this reduced system: a momentum constraint $C_\theta$ that generates rigid rotations in $\theta$, and a Hamiltonian constraint $C_\text{G}$ that generates time reparametrizations. The latter can be split into two terms, $C_\text{G}=C_\text{BI}+C_\text{inh}$, where the homogeneous term $C_\text{BI}$ denotes the Hamiltonian constraint of the Bianchi I model with LRS and a homogeneous massless scalar, and where $C_\text{inh}$ rules the dynamics of the inhomogeneities (non-zero Fourier modes), coupling them with the homogeneous sector.

The hybrid approach \cite{hybrid-matter} consists in adopting a Fock quantization for both the gravitational and matter inhomogeneities, a standard Schr\"odinger representation for the homogeneous massless scalar field $\phi$, and a loop quantization for the Bianchi I variables \cite{MMP} within the so-called improved dynamics scheme \cite{awe1}. For the loop quantization of the homogeneous gravitational sector (Bianchi I phase space with LRS, which is four dimensional), one introduces some particular real variables. Adopting the conventions and notations of \cite{hybrid-approx}, we call these variables $\{v,b,\lambda_\theta,b_\theta\}$. They have the following Poisson brackets 
\begin{align}\label{poisson}
\{b,v\}&=\frac2{\hbar}, \quad \{b_\theta,\lambda_\theta\}=\frac2{\hbar}\frac{\lambda_\theta}{v}, \qquad \{\lambda_\theta,v\}=0,\nonumber\\\{b_\theta,v\}&=\frac2{\hbar}, \quad \{b_\theta,b\}=\frac2{\hbar v} (b_\theta-b),\quad \{\lambda_\theta,b\}=0.
\end{align}
The absolute value of $v$ is proportional to the physical volume of the Bianchi I universe (which has compact topology), while $\lambda_\theta$ measures the anisotropy. The symbol $\hbar$ denotes the reduced Planck constant

The above set of variables is not canonical but it proves to be convenient for the loop quantization. In this quantization, $\hat{v}$ and $\hat{\lambda}_\theta$ act as multiplicative operators. Their eigenstates $|v,\lambda_\theta\rangle$ (with $v,\lambda_\theta\in\mathbb{R}$) provide a basis for the Hilbert space of the homogeneous gravitational sector, the inner product being the discrete one: $\langle v',\lambda'_\theta|v,\lambda_\theta\rangle=\delta_{v',v}\delta_{\lambda'_\theta,\lambda_\theta}$. Because of this,  $\hat{v}$ and $\hat{\lambda}_\theta$ have discrete spectra that run over the whole real line. As a consequence,  there are no well defined operators representing the variables $b_a$ (here, $b_{a}$ stands both for $b_\theta$ and $b$), but instead one can define operators $\widehat{e^{\pm ib_{a}}}$ for their ``holonomy'' elements.
Then, one represents the canonical commutation relations $[ \widehat{e^{\pm ib_{a}}},\widehat{v}]=i\hbar\widehat{\{e^{\pm ib_{a}},v}\}$ and $[ \widehat{e^{\pm ib_{a}}},\widehat{\lambda_\theta}]=i\hbar\widehat{\{e^{\pm ib_{a}},\lambda_\theta}\}$. Since the Poisson brackets of the $b_a$'s with $v$ are constant, these holonomies produce a constant shift on the volume variable. However, $\widehat{e^{\pm ib_{\theta}}}$ produces a state-dependent shift on $\lambda_\theta$. Explicitly,
\begin{align}\label{hol}
\widehat{e^{\pm ib}} |v,\lambda_\theta\rangle=|v\pm2,\lambda_\theta\rangle,\qquad  \widehat{e^{\pm ib_\theta}} |v,\lambda_\theta\rangle=\left|v\pm2, \lambda_\theta\pm\frac{2\lambda_\theta}{v}\right\rangle.
\end{align}
Therefore, $\widehat{e^{\pm ib}}$ and $\widehat{e^{\pm ib_{\theta}}}$ do not commute.

Using these operators, and adopting a standard Schr\"odinger representation for $\phi$, so that its momentum is promoted to the operator $\hat{p}_\phi=-i\hbar \partial_\phi$ on the usual Hilbert space $L^2(\mathbb{R},d\phi)$, the quantum operator representing the Bianchi I term $C_\text{BI}$ reads
\begin{align}
\hat{C}_{\text{BI}}&= \hat{C}_{\text{FRW}}-\frac{\pi G\hbar^2}{8}(\hat{\Omega}\hat{\Theta}+\hat{\Theta}\hat{\Omega})\quad;\quad  \hat{C}_{\text{FRW}}=-\frac{3\pi G\hbar^2}{8}\hat{\Omega}^2+\frac{\hat{p}_\phi^2}{2}.
\end{align}
Here, $G$ is the Newton constant, and $\hat{\Theta}\equiv \hat{\Theta}_\theta-\hat{\Omega}$, with $\hat{\Omega}$ and $\hat{\Theta}_\theta$ two symmetrized operators that represent, respectively,  $2v\sin(b)$ and $2v\sin(b_\theta)$:
\begin{align}\label{Omegaop}
\hat{\Omega}&=\sqrt{|\hat v|}\left[\widehat{{\rm sign}(v)}\widehat{\sin(b)}+\widehat{\sin(b)}\widehat{{\rm sign}(v)}\right]\sqrt{|\hat v|},\\
\hat{\Theta}_\theta&=\sqrt{|\hat v|}\left[\widehat{{\rm sign}(v)}\widehat{\sin(b_\theta)}+\widehat{\sin(b_\theta)}\widehat{{\rm sign}(v)}\right]\sqrt{|\hat v|}. \label{Thetaop}
\end{align}
Thus, the LRS Bianchi I constraint is formed by that of the flat FRW model coupled to a massless scalar, $\hat{C}_{\text{FRW}}$, plus a term accounting for the anisotropies. 

Thanks to the factor ordering chosen for $\hat{\Omega}$ and $\hat{\Theta}_\theta$ \cite{MMP,mmo}, $\hat{C}_{\text{BI}}$ decouples the states $|v,\lambda_\theta\rangle$ with $v=0$ and/or $\lambda_\theta=0$ from the rest of the basis states, so that we can remove them in the rest of our considerations. These states with $v=0$ and/or $\lambda_\theta=0$ are the quantum analogs of the cosmological singularities, so that their decoupling implies a kinematical resolution of them. The action of the operator $\hat{C}_{\text{BI}}$ also decouples states with positive $v$ from states with negative $v$, and likewise for the variable $\lambda_\theta$. Thanks to this property one can restrict the study e.g. to states with strictly positive $v$ and $\lambda_\theta$. For convenience, we then introduce the definition $\Lambda\equiv \ln(\lambda_\theta)$ and relabel the basis states for the homogeneous gravitational sector as $|v,\Lambda\rangle$ (with $v\in\mathbb{R}^+$, $\Lambda\in\mathbb{R}$).

On the other hand, for the inhomogeneous sector one adopts a Fock quantization, expressing the non-zero modes of the fields in terms of annihilation and creation-like variables, that are later promoted to operators acting on $n$-particle states, which in turn provide a basis for the Fock space. The Fock quantization adopted is uniquely determined, up to unitary transformations, by the conditions of invariance of the vacuum under rigid rotations in $\theta$ and a unitary quantum dynamics \cite{Fock}. These criteria, in particular, require the scaling of the gravitational waves and the scalar field by a specific homogeneous factor. The unitary class of these invariant (under rigid rotations) Fock representations contain the ``massless'' representation, in which annihilation and creation-like variables are constructed ignoring all mass terms in the frequency of the modes. We refer to \cite{hybrid-matter, hybrid-approx} for all these details. In the following, we denote the annihilation operator associated with the mode $m\in \mathbb{Z}-\{0\}$ in the adopted massless representation by $\hat{a}^{(\alpha)}_m$, where $\alpha=\xi$ denotes the gravitational wave and $\alpha=\varphi$ denotes the matter field, both conveniently rescaled.  On the other hand, the $n$-particle states of the Fock space are called $|\mathfrak{n}^\xi,\mathfrak{n}^\varphi\rangle$, where we are using the notation $\mathfrak{n}^\alpha=\{\cdots,n^\alpha_{-m},\cdots,n^\alpha_{-1},n^\alpha_1,\cdots,n^\alpha_m,\cdots\}$.
As a result,  $C_\text{inh}$ is promoted to the operator
\begin{align}\label{constra}
\hat{C}_\text{inh}=\frac{2\pi G \hbar^{2}}{\beta} \widehat{e^{2\Lambda}}\hat{H}_0+ \frac{\pi G \hbar^2 \beta}{4}  \widehat{e^{-2\Lambda}}\hat{D}\hat{\Omega}^2\hat{D}\hat{H}_\text{I}.
\end{align}
Here, $\beta$ is a constant that depends on some parameters of the loop quantization [namely, $\beta=[G \hbar/(16\pi^{2} \gamma^{2}\Delta)]^{1/3}$, $\gamma$ being the Immirzi parameter \cite{Immirzi} and $\Delta$  the gap in the spectrum of area eigenvalues allowed in loop quantum gravity], $\hat{H}_0$ is the free contribution of the non-zero modes,
\begin{align}
\hat{H}_0=\sum_{\alpha\in{\xi,\varphi}}\,\sum_{m\in\mathbb{Z}-\{0\}} |m| \, \hat{a}^{(\alpha)\dagger}_m \hat{a}^{(\alpha)}_m,
\end{align}
$\hat{H}_\text{I}$ is a self-interaction term for these inhomogeneities
\begin{align}
\hat{H}_\text{I}=\sum_{\alpha\in{\xi,\varphi}}\,\sum_{m\in\mathbb{Z}-\{0\}}\frac1{2|m|} \left(2 \hat{a}^{(\alpha)\dagger}_m \hat{a}^{(\alpha)}_m+\hat{a}^{(\alpha)\dagger}_m \hat{a}^{(\alpha)\dagger}_{-m}+\hat{a}^{(\alpha)}_m \hat{a}^{(\alpha)}_{-m}\right),
\end{align}
and $\hat{D}$ represents the product of the volume by its inverse, which is regularized in a well established way in LQC, so that one gets
\begin{align}
\hat{D}|v\rangle=D(v)|v\rangle,\qquad D(v)\equiv v\left(\sqrt{v+1}-\sqrt{|v-1|}\right)^2.
\end{align}

The total constraint operator, $\hat{C}_\text{G}=\hat{C}_{\text{BI}}+\hat{C}_\text{inh}$, is densely defined on the Hilbert space spanned by the states $|v,\Lambda,\phi,\mathfrak{n}^\xi,\mathfrak{n}^\varphi\rangle$, with inner product
\begin{align}
\langle v',\Lambda',\phi',\mathfrak{n'}^\xi,\mathfrak{n'}^\varphi|v,\Lambda,\phi,\mathfrak{n}^\xi,\mathfrak{n}^\varphi\rangle=\delta_{v',v}\delta_{\Lambda',\Lambda}\delta(\phi'-\phi)\delta_{\mathfrak{n'}^\xi,\mathfrak{n}^\xi}\delta_{\mathfrak{n'}^\varphi,\mathfrak{n}^\varphi}.
\end{align}
Here, all the $\delta$'s are Kronecker deltas, except $\delta(\phi'-\phi)$ which is the Dirac delta, since $\phi$ is a continuous variable. This Hilbert space is non-separable, as long as the states $|v,\Lambda\rangle$ that span the homogeneous gravitational sector form a non-countable basis (we recall that $v\in\mathbb{R}^+$ but this set has to be understood with discrete topology, and likewise for $\Lambda\in\mathbb{R}$). However, the operators  $\hat{\Omega}$ and $\hat{\Theta}_\theta$, which are the only ones acting non-diagonally on the states  $|v,\Lambda\rangle$, preserve subspaces that provide separable superselection sectors, and hence the same happens for the full operator $\hat{C}_\text{G}$. Indeed, we can restrict the domain of definition of $\hat{C}_\text{G}$ to states with label $v$ belonging to any of the semilattices of step four
\begin{align}
\mathcal
L_{\varepsilon}^+=\{\varepsilon+4k;\;k\in\mathbb{N}\},\qquad\varepsilon\in(0,4],
\end{align}
each of them characterized by a minimum (strictly positive) eigenvalue $\varepsilon$ for $v$. On the other hand, in what concerns the anisotropy variable $\Lambda$, we get that the iterative action of the constraint operator relates any given state $|v,\Lambda^\star\rangle$ only with states whose quantum number $\Lambda$ is of the form 
$\Lambda=\Lambda^\star+\omega_\varepsilon$, with $\omega_\varepsilon$
belonging to the set $\mathcal{W}_\varepsilon$ defined as \cite{hybrid3}
\begin{equation}\label{W-set}
\left\{z\,\ln\left(\frac{\varepsilon-2}{\varepsilon}\right)+\sum_{m,
	n\in\mathbb{N}}{k_n^m}\ln\left(\frac{\varepsilon+2m}{\varepsilon+2n}\right);\;
k_n^m\in\mathbb{N},\; z\in\mathbb{Z}\text{ if } \varepsilon>2,\;z=0\text{ if }
\varepsilon\le2\right\}.
\end{equation}
This set is countable and dense in the real line.

To conclude this section, let us write the momentum constraint operator, given by \cite{hybrid-matter}
\begin{align}\label{dif}
\hat{C}_\theta=\sum_{\alpha\in{\xi,\varphi}}\,\sum_{m\in\mathbb{N}^{+}} m\left( \hat{a}^{(\alpha)\dagger}_m \hat{a}^{(\alpha)}_m - \hat{a}^{(\alpha)\dagger}_{-m} \hat{a}^{(\alpha)}_{-m}\right).
\end{align}
It only acts on the inhomogeneous sector, and requires that the total field momentum vanish by restricting the numbers of particles so that
\begin{align}\label{mom}
\sum_{m\in\mathbb{N}^{+}}m\left(n^\xi_m+n^\varphi_m-n^\xi_{-m}-n^\varphi_{-m}\right)=0.
\end{align}

\subsection{Approximate solutions}
\label{iib}

In \cite{hybrid-approx} it was proven that certain solutions to the constraint
\begin{align}\label{const-app}
\hat{C}_{\text{app}}=\hat{C}_{\text{FRW}}+\frac{2\pi G \hbar^{2}}{\beta} \widehat{e^{2\Lambda}}\hat{H}_0=-\frac{3\pi G\hbar^2}{8}\hat{\Omega}^2+\frac{\hat{p}_\phi^2}{2}+\frac{2\pi G \hbar^{2}}{\beta}\widehat{e^{2\Lambda}}\hat{H}_0
\end{align}
are approximate solutions to the full Gowdy constraint $\hat{C}_\text{G}=\hat{C}_{\text{BI}}+\hat{C}_\text{inh}$, because the action of both the anisotropy term and of the term involving the self-interaction of the inhomogeneities on those states can be disregarded.

\begin{figure}[ht]
	\includegraphics[width=0.48\textwidth]{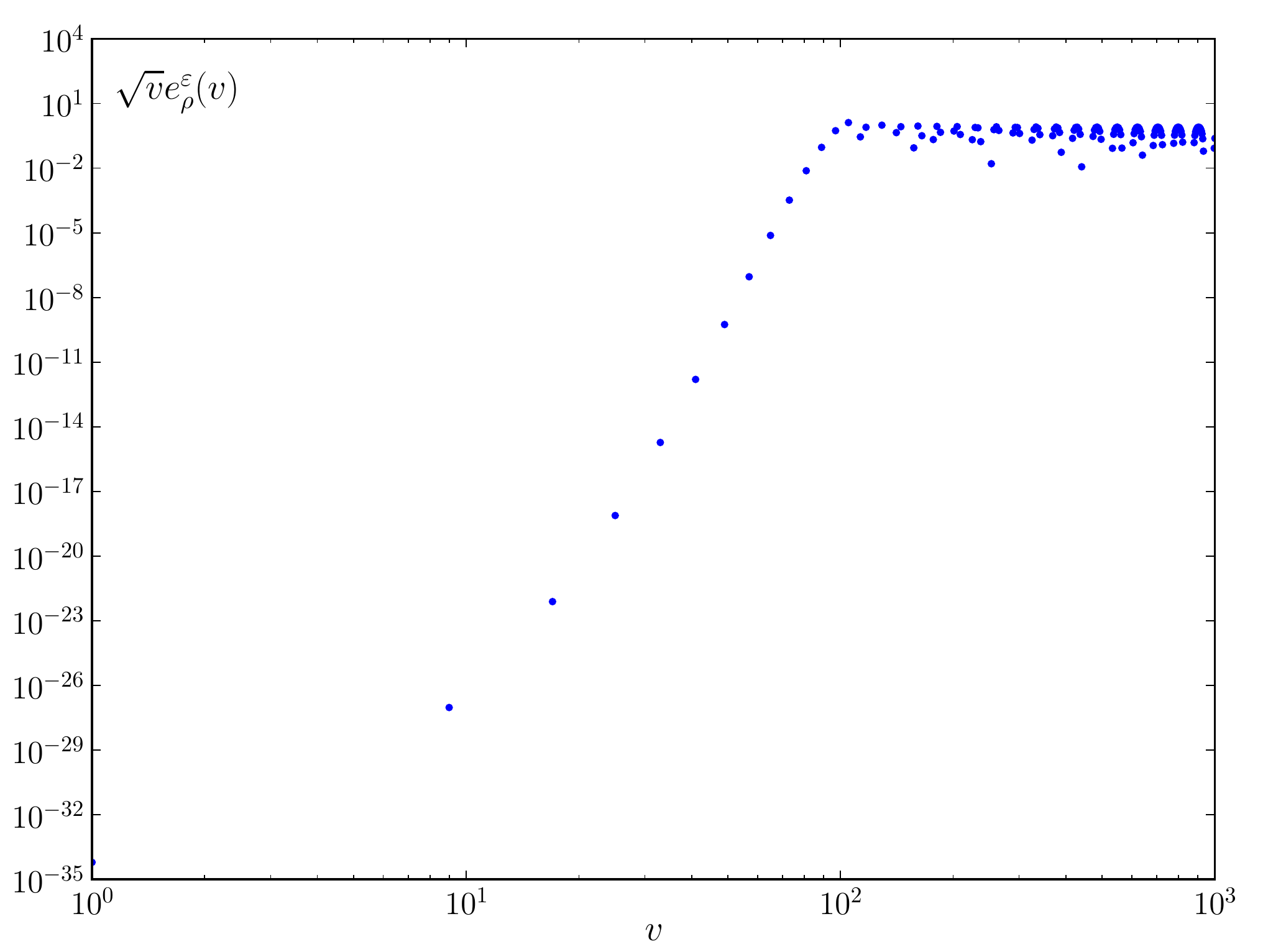}
	\includegraphics[width=0.48\textwidth]{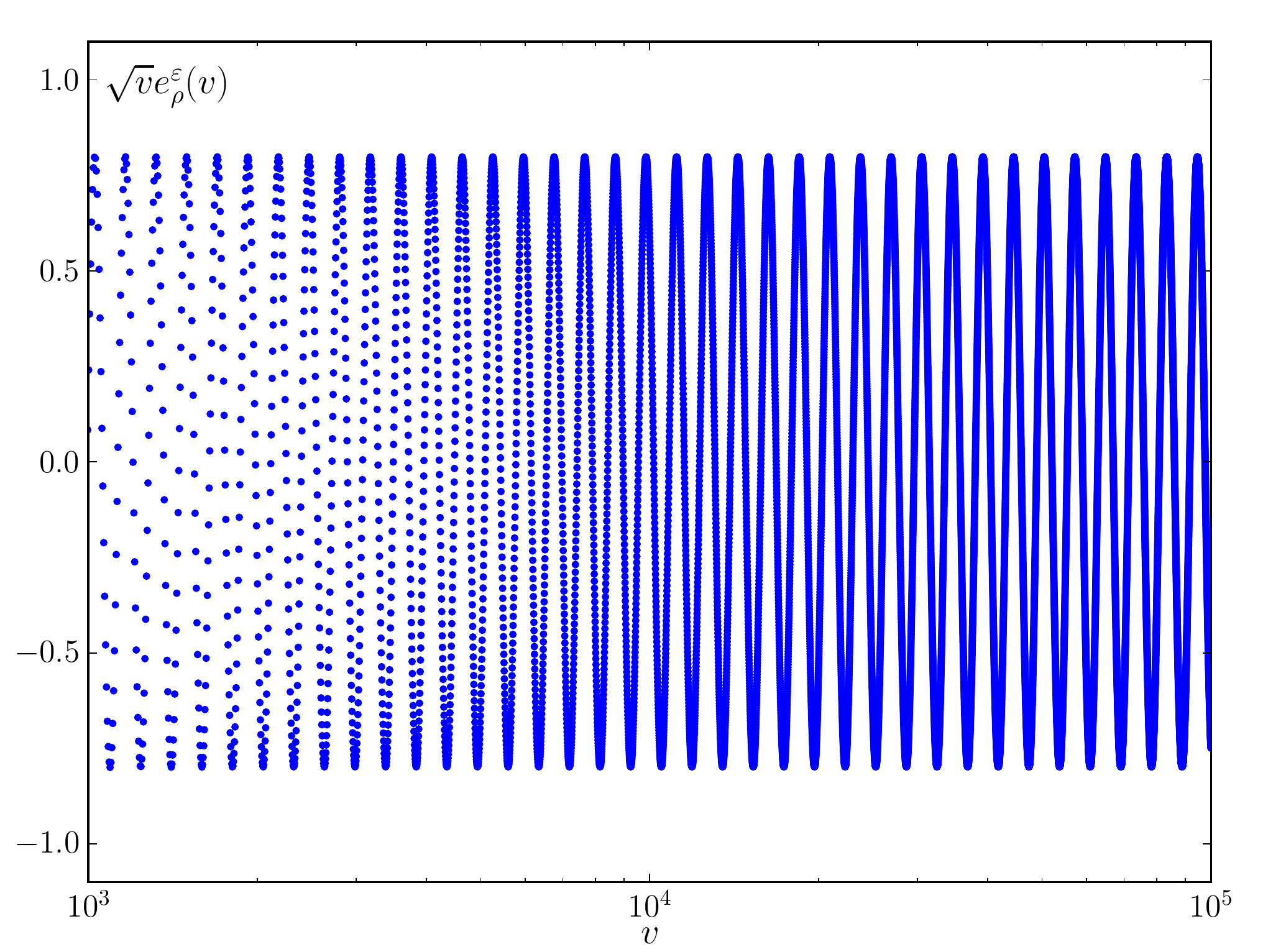}
	\caption{The function  $\sqrt{v}e^\varepsilon_\rho(v)$ in the case $\varepsilon=1$ and $\rho=200$. The left graph displays the exponential suppression of the function for small values of $v$. The right graph shows the oscillatory behavior at large $v$. In both graphs, a logarithmic scale is used in the horizontal axis.}
	\label{fig:FRWaut}
\end{figure}

The first key point to obtain those solutions relies on the behavior of the eigenstates of the FRW operator $\hat{\Omega}^2$. This operator is essentially self-adjoint, with absolutely continuous, non-degenerate, and positive spectrum \cite{mmo}. In each of the superselection sectors spanned by states $|v\rangle$ with support on the semilattices $\mathcal{L}^+_\varepsilon$, the delta-normalized eigenstates of  $\hat{\Omega}^2$  with eigenvalue $\rho^2$, $|e^\varepsilon_\rho\rangle=\sum_{v\in\mathcal{L}^+_\varepsilon}e^\varepsilon_\rho(v)|v\rangle$, provide a resolution of the identity: $\mathbb{I}=\int_{0}^{\infty} d\rho |e^\varepsilon_{ \rho}\rangle \langle e^\varepsilon_\rho|$. The eigenfunctions can be chosen to be real and present an essential property for the approximations to hold, namely that, for $\rho \gg 10$, $e^\varepsilon_\rho(v)$ is exponentially suppressed for $v\lesssim \rho/2$. On the other hand, for $v\gg \rho/2$, these eigenfunctions display an oscillatory behavior. See Fig. \ref{fig:FRWaut}, which has been extracted from \cite{hybrid-approx}. The exponential suppression of the region  $v\lesssim \rho/2$ is a characteristic feature of the quantum geometry in the formalism of LQC. This phenomenon is behind the occurrence of a quantum bounce that cures the cosmological singularities in the loop quantization \cite{mmo}. In the vicinity of this bounce, quantum geometry effects change drastically the behavior of the gravitational interaction with respect to Einstein's theory, invalidating the expectations based on general relativity \cite{effective}. This change has been investigated and confirmed as well in the presence of anisotropies \cite{param} and of inhomogenities \cite{tarrio}.

Thanks to the mentioned suppression for small values of $v$, one can check that the inverse volume corrections play no role as long as $\rho\gg 10$, namely we have $\hat{D}\hat{\Omega}^2\hat{D} |e^\varepsilon_\rho\rangle \simeq \hat{\Omega}^2 |e^\varepsilon_\rho\rangle$, and therefore
\begin{align}\label{approxD}
\hat{C}_\text{I}= \frac{\pi G \hbar^2 \beta}{4}  \widehat{e^{-2\Lambda}}\hat{D}\hat{\Omega}^2\hat{D}\hat{H}_\text{I}\simeq  \frac{\pi G \hbar^2 \beta}{4}  \widehat{e^{-2\Lambda}}\hat{\Omega}^2\hat{H}_\text{I},
\end{align}
when acting on states in the sector $\rho\gg 10$. This is the first approximation introduced in \cite{hybrid-approx}.

The next step is to approximate the anisotropy operator $\hat{\Omega}\hat{\Theta}+\hat{\Theta}\hat{\Omega}$. This is a complicated operator, which does not commute with the FRW operator $\hat{\Omega}^2$, and which does not split into the product of two operators, one acting on the volume variable $v$, and another acting on the anisotropy variable $\Lambda$. Nonetheless, as it was shown in \cite{hybrid-approx}, it can be approximated by an operator that does factorize if one considers only its action on states $|e_{\rho}^{\varepsilon}\rangle \otimes | f(\Lambda)\rangle$, with $|f(\Lambda)\rangle=\sum_{\omega_\varepsilon\in\mathcal{W}_\varepsilon}f(\Lambda^\star+\omega_\varepsilon)|\Lambda^\star+\omega_\varepsilon\rangle$ and such that the profile $f(\Lambda)$ can be extended to a smooth function in the real line that, for variations in $\Lambda$ smaller than a certain scale $q_{\varepsilon}$ (depending on the considered superselection sector), satisfies:
\begin{equation}\label{tayl}
f(\Lambda+\Lambda_{0})\simeq f(\Lambda) +\Lambda_{0}\partial_{\Lambda} f(\Lambda).
\end{equation}
Thus, for $\Lambda_0\leq q_{\varepsilon}$, the approximation disregards higher-order terms in $\partial_\Lambda$ in a Taylor expansion. On these states, then, the anisotropy term is approximated by a factorizing operator: $\hat{\Omega}\hat{\Theta}+\hat{\Theta}\hat{\Omega}\simeq -2\hat{\tilde\Omega}\hat{\Theta}' $. Here,
\begin{align}\label{omegatilde}
\hat{\tilde\Omega}=\sqrt{\frac{|\hat v|}{2}}\left[\widehat{{\rm sign}(v)}\widehat{\sin(2b)}+\widehat{\sin(2b)}\widehat{{\rm sign}(v)}\right]
\sqrt{\frac{|\hat v|}{2}}
\end{align}
is completely analogous to the geometry operator $\hat{\Omega}$, introduced in \eqref{Omegaop}, except for the replacement of the canonically conjugate variables $(v,b)$ with the pair $(v/2,2b)$. On the other hand
\begin{align}\label{zetaprim}
\hat{\Theta}'|\Lambda\rangle= i\frac2{q_\varepsilon}\left(|\Lambda+q_\varepsilon\rangle-|\Lambda-q_\varepsilon\rangle\right),
\end{align}
where one chooses $q_\varepsilon\in\mathcal{W}_\varepsilon$  so that $\hat{\tilde\Omega}\hat{\Theta}'$ preserves the  superselection sectors of  $\hat{\Omega}\hat{\Theta}+\hat{\Theta}\hat{\Omega}$. Furthermore, $\hat{\Theta}'$ preserves the lattices of constant step $\mathcal{L}^{q_\varepsilon}_{\Lambda^\star} =\{ \Lambda^\star+nq_\varepsilon;\, n\in\mathbb{Z}\}$. Therefore, as far as the anisotropy variable is concerned, we can further restrict the study to the subspace spanned by the states $|\Lambda\rangle$ with support in any of those lattices. In the considered regime with $\rho\geq \rho^{\star}\gg 10$, $q_\varepsilon$ is defined as 
\begin{align}\label{scale}
q_\varepsilon=\ln\left(1+\frac{2}{v^\star}\right),\qquad v^\star={\rm max}\left\{v\in\mathcal{L}^+_\varepsilon \text{ such that } v<  \frac{\rho^\star}{2} \right\}.
\end{align}
The above approximation for the anisotropy term is specially relevant for the study of the Bianchi I model, $\hat{C}_{\text{BI}}\simeq\hat{C}_{\text{FRW}}+\pi G\hbar^2\hat{\tilde\Omega}\hat{\Theta}'/4$, as the spectral properties of $\hat{\Theta}'$ are completely known \cite{hybrid-approx}, in contrast to the situation with the original operator, which is unmanageable. 

As we have already said, in \cite{hybrid-approx} states were found on which both the action of the anisotropy term and of $\hat{C}_\text{I}$ can be disregarded. This is possible by considering states with the following anisotropy Gaussian profiles
\begin{align}\label{ani}
f(\Lambda)= \frac{\sqrt{\sigma_{s}}}{\sqrt[4]{\pi}} e^{-\frac{\sigma^{2}_{s}}{2q_{\epsilon}^{2}}(\Lambda-\bar{\Lambda})^{2}}.
\end{align}
Indeed, the action of the anisotropy term is negligible in comparison with that of the FRW operator for $\sigma_{s}\ll\pi/2$; besides, the action of $\hat{C}_\text{I}$ can be ignored as long as $\bar\Lambda\gg 1 \gg q_{\varepsilon}^2/\sigma^{2}_{s}$ and the content of inhomogeneities is sufficiently small, so that the exponential suppression produced by $\widehat{e^{-2\Lambda}}$ kills the contribution of $\hat{H}_\text{I}$.
In summary, for states with the above anisotropy profiles, the Hamiltonian constraint of the Gowdy model can be approximated by the constraint operator \eqref{const-app}. 

Before continuing our discussion, it is worth clarifying that, in spite of the dynamical behavior proven for these states with respect to the constraint of the system (namely, a homogeneous and isotropic one), their Gaussian profiles are not really peaked on isotropic trajectories, but rather on very anisotropic ones. Actually, isotropic trajectories satisfy the relation $3\Lambda= \ln{(v/2)}$, which reflects the fact that $\lambda_{\theta}=e^{\Lambda}$ coincides in isotropic settings with the geometrical average scale factor of the model, $v^{1/3}$, up to proportionality factors \cite{hybrid-approx}. The Gaussian wave functions \eqref{ani}, however, are peaked on trajectories with constant value of $\Lambda$, given by $\bar{\Lambda}$. These trajectories do not correspond to isotropic and homogeneous solutions of the classical Gowdy model in general reativity. From this perspective, the fact that these Gaussian profiles can ultimately be associated with states that are approximate solutions of an FRW Hamiltonian constraint must be traced to the collective behavior of the anisotropies and inhomogeneities when the exact quantum constraint of the Gowdy model is considered.

The approximated constraint is easy to impose on the states under consideration because  $\hat{C}_{\text{app}}$ can be readily diagonalized. 
The resulting solutions to the constraint 
\begin{align}\label{constrainteq}
(\Psi |\hat{C}_{\text{app}}^{\dagger}|\phi,v,\Lambda,\mathfrak{n}^{\xi},\mathfrak{n}^{\phi}\rangle=0
\end{align}
are states
\begin{equation}\label{gen-states}
(\Psi|=\int_{-\infty}^{\infty} d{\phi}\sum_{v\in\mathcal{L}^+_{\varepsilon}}\sum_{\Lambda\in\mathcal{L}^{q_\varepsilon}_{\Lambda^\star}}
\sum_{\mathfrak{n}^{\xi},\mathfrak{n}^{\varphi}}\Psi(\phi,v,\Lambda,\mathfrak{n}^{\xi},\mathfrak{n}^{\varphi})
\langle{\phi},v,\Lambda,\mathfrak{n}^{\xi},\mathfrak{n}^{\phi}|
\end{equation} with profiles of the form
\begin{align}\label{profiles}
\Psi(\phi,v,\Lambda,\mathfrak{n}^{\xi},\mathfrak{n}^{\varphi})&= \int_{-\infty}^{\infty} dp_{\phi}\, \psi(p_\phi)\chi(\mathfrak{n}^{\xi},\mathfrak{n}^{\varphi}) f(\Lambda)e^{\varepsilon}_{\rho(p_\phi,\bar\Lambda,\mathfrak{n}^{\xi},\mathfrak{n}^{\varphi})}(v) e_{p_\phi}(\phi),
\end{align}
where $
e_{p_\phi}(\phi)=e^{\frac{i}{\hbar}p_\phi\phi}/\sqrt{2\pi\hbar}$ are the usual plane-waves that diagonalize $\hat{p}_\phi^2=-\hbar^2\partial^2_\phi$,
\begin{align}
\rho(p_\phi,\Lambda,\mathfrak{n}^{\xi},\mathfrak{n}^{\varphi})&=\sqrt{\frac{4}{3\pi G\hbar^2}p_\phi^{2}+\frac{16}{3\beta} e^{2\Lambda}H_{0}(\mathfrak{n}^{\xi},\mathfrak{n}^{\varphi})},\\
H_{0}(\mathfrak{n}^{\xi},\mathfrak{n}^{\varphi})&=\sum_{m\in\mathbb{Z}-\{0\}}\!\!\!|m|(n^\xi_m+n^\varphi_m),
\end{align}
and $f(\Lambda)$ is given in \eqref{ani}. On the other hand, the wave function of $n$-particles states, $\chi(\mathfrak{n}^{\xi},\mathfrak{n}^{\varphi})$, should be chosen in such a way that the content of inhomogeneities is small, so that the approximation of disregarding $\hat{C}_\text{I}$ holds, and such that the momentum constraint \eqref{mom} is satisfied. Concerning the function $\psi(p_\phi)\in L^2(\mathbb{R},d\phi)$, it can be chosen to be e.g. a function of compact support on a region of sufficiently large values of $p_\phi$, for instance $p_\phi \geq p_\phi^\star\gg \sqrt{75\pi G}\hbar$, to make sure that we are in the sector $\rho\geq\rho^\star\gg 10$ under consideration. Here, we are defining $p_\phi^\star$ through the relation
\begin{align}
\rho^\star=\frac{2 |{p}^\star_\phi|}{\sqrt{3\pi G\hbar^2}} .
\end{align}

Let us emphasize that the above solutions verify the effective constraint
\begin{align}\label{eff-cons}
\rho=\frac{2 |\tilde{p}_\phi|}{\sqrt{3\pi G\hbar^2}} , \quad |\tilde p_\phi|\equiv \sqrt{p_\phi^{2}+\frac{4\pi G \hbar^2}{\beta}e^{2\bar\Lambda}H_0},
\end{align}
which corresponds to that of a flat FRW model with a (redefined) massless scalar $\tilde{p}_\phi$, since both $\bar\Lambda$ and $H_0$ are constants, like the conserved quantity $p_\phi$. Namely, the inhomogeneous and anisotropic contribution of these states behave in the Hamiltonian constraint as an effective homogeneous massless scalar field. This shows that anisotropic and inhomogeneous states may lead indeed to isotropic dynamical effective descriptions.

An observation not done in \cite{hybrid-approx} is in order. We note that we do not need the massless scalar $\Phi$ to obtain the above behavior: even in the case of the Gowdy model in vacuo, the inhomogeneities due solely to the gravitational wave $\xi$ lead to an effective isotropic massless scalar field. In that case, to ensure that the solutions correspond to the regime $\rho\gg 10$ where the approximations hold, we would just need to require
\begin{align}
e^{\bar\Lambda}\sqrt{\frac{16H_0(n^\xi)}{3\beta}}\gg10,
\end{align}
which is compatible with the condition $\bar\Lambda\gg 1\gg  {q_{\varepsilon}^2}/\sigma^{2}_{s}$ that the peak of the Gaussian has to satisfy.

\section{New class of approximate solutions: Mimicking perfect fluid behavior}
\label{sec:mimic}

We now proceed to construct other approximate solutions to the Gowdy model that are still exact solutions to the constraint $\hat{C}_\text{app}$ \eqref{const-app}, but for which the term that goes with $\hat{H}_0$ effectively depends on $v$, so that this term does not longer provide a constant. As we will show, this new behavior can be attained by generalizing the previously considered states, which have a homogeneous gravitational contribution given by $|e_{\rho}^{\varepsilon}\rangle \otimes | f(\Lambda)\rangle$ with anisotropy profile \eqref{ani}, to states in which the homogeneous gravitational part is of the form $| g(v,\Lambda)\rangle=\sum_{v\in\mathcal{L}^+_{\varepsilon}}\sum_{\Lambda\in\mathcal{L}^{q_\varepsilon}_{\Lambda^\star}}g(v,\Lambda)|v,\Lambda\rangle$
with 
\begin{align}\label{ani-v}
g(v,\Lambda)=N(v) f(v,\Lambda),\qquad f(v,\Lambda)= e^{-\frac{\sigma^{2}_{s}}{2q_{\epsilon}^{2}}[\Lambda-\bar{\Lambda}(v)]^{2}}.
\end{align}
In particular, the variation with $v$ of the peak $\bar{\Lambda}(v)$ of the Gaussian profile makes it possible that the state concentrates on isotropic trajectories, case which requires the anisotropy variable to be equal to $\ln{(v/2)}/3$, as we commented above.

The dependence on $v$ of the above function $f(v,\Lambda)$ implies that $\hat{C}_\text{app}$ cannot be straightforwardly diagonalized anymore when acting on states with this type of homogeneous gravitational part; however, solutions can still be constructed. 

Let us recall that when $\bar\Lambda$ is  a constant, owing to the behavior of the eigenfunctions $e^{\varepsilon}_{\rho}(v)$ of the FRW operator, the solutions are highly suppressed for $v\leq v^\star$, with $v^\star$ defined in \eqref{scale}, provided that the contributing eigenvalues $\rho$ are not smaller than $\rho^{\star}\gg10$.  In the present case with $\bar\Lambda(v)$ the solutions also need to display a similar behavior, for the approximations carried out in \cite{hybrid-approx} to still hold, namely the function $N(v)$ has to be highly suppressed for $v\leq v_m$, for a certain value $v_m\gg 10$. 
For the moment being, let us assume that the above condition is true, and let us check under which new conditions the approximations of \cite{hybrid-approx} extend to the present case. Then, in Sec. \ref{secB} we will explain how to construct solutions with the desired properties. 

\subsection{Approximating the anisotropy and interaction terms}
\label{iiia}

We first note that, if there are no contributions with $v\leq v_m$, as we are assuming, then approximation \eqref{approxD} is still verified, because $D(v)$ is very close to 1 for the relevant values of $v$. 

In order to deal with the anisotropy operator, we consider states $|g(v,\Lambda)\rangle$ such that $g(v,\Lambda)$ can be extended to a smooth function in the real line of $\Lambda$ (for all $v$), so that 
\begin{equation}\label{tayl2}
g(v,\Lambda+\Lambda_{0})\simeq g(v,\Lambda) +\Lambda_{0}\partial_{\Lambda} g(v,\Lambda) 
\end{equation}
for displacements $\Lambda_{0}\leq q_{\varepsilon}$, where now this scale is defined as
\begin{align}\label{scale2}
q_\varepsilon=\ln\left(1+\frac{2}{v_m}\right).
\end{align}
For such states, the approximation $\hat{\Omega}\hat{\Theta}+\hat{\Theta}\hat{\Omega}\simeq -2\hat{\tilde\Omega}\hat{\Theta}'$ is valid. The proof follows as in \cite{hybrid-approx}. We first compute the action of the operator  $\hat{\Omega}\hat{\Theta}+\hat{\Theta}\hat{\Omega}$ on states $|g(v,\Lambda)\rangle$:
\begin{align}\label{action-ani}
\langle {v}^{\prime},&{\Lambda}^{\prime}|\hat{\Omega}\hat{\Theta}+\hat{\Theta}\hat{\Omega}|g(v,\Lambda)\rangle   \nonumber \\
& = -y_{--}({v}^{\prime})\left[g({v}^{\prime}-4,{\Lambda}^{\prime}+d_{{v}^{\prime}}(-4)-d_{{v}^{\prime}}(-2))-2g({v}^{\prime}-4,{\Lambda}^{\prime})+g({v}^{\prime}-4,{\Lambda}^{\prime}+d_{{v}^{\prime}}(-2))\right] \nonumber\\
& \quad - y_{++}({v}^{\prime})\left[g({v}^{\prime}+4,{\Lambda}^{\prime}+d_{{v}^{\prime}}(4)-d_{{v}^{\prime}}(2))-2g({v}^{\prime}+4,{\Lambda}^{\prime})+g({v}^{\prime}+4,{\Lambda}^{\prime}+d_{{v}^{\prime}}(2))\right] \nonumber \\
&  \quad +   y_{+-}({v}^{\prime})\left[g({v}^{\prime},{\Lambda}^{\prime}+d_{{v}^{\prime}}(-2))-2g({v}^{\prime},{\Lambda}^{\prime})+g({v}^{\prime},{\Lambda}^{\prime}-d_{{v}^{\prime}}(-2))\right] \nonumber\\
& \quad +  y_{-+}({v}^{\prime})\left[g({v}^{\prime},{\Lambda}^{\prime}+d_{{v}^{\prime}}(2))-2g({v}^{\prime},{\Lambda}^{\prime})+g({v}^{\prime},{\Lambda}^{\prime}-d_{{v}^{\prime}}(2))\right] .
\end{align}
Here, we have introduced the notation $\langle {v}^{\prime},{\Lambda}^{\prime}|$ for bra-states to make clear that the computed operator elements are generic, and are not limited to the diagonal ones in the $| v,\Lambda\rangle$-basis. In addition, we have defined
\begin{align}\label{def2}
y_\pm(v)=\frac{1+\text{sign}(v\pm2)}{2}\sqrt{v(v\pm2)},
\end{align}
\begin{align}\label{defini2}
y_{\pm\pm}(v)=y_\pm(v) y_\pm(v\pm2),\qquad y_{\mp\pm}(v)=y_\pm(v) y_\mp(v\pm2),
\end{align}
and
\begin{align}\label{def3}
d_v(n)\equiv \ln\left(1+\frac{n}{v}\right).
\end{align}
To derive \eqref{action-ani}, we have taken into account that a shift on the label of a volume eigenstate translates into the opposite shift on the wave functions in a $v$ representation. In addition, we have used the identities  
\begin{eqnarray}y_{++}(v)&=y_{--}(v+4),\quad\quad\quad\quad\quad\quad\quad y_{--}(v)&=y_{++}(v-4);\\ d_v(\pm 2)&=d_{v\pm 4}(\mp 2)-d_{v\pm 4}(\mp 4), \quad\quad d_{v\pm 4}(\mp 2)&=d_{v}(\pm 2)-d_{v}(\pm 4).
\end{eqnarray}

Now, for smooth functions satisfying \eqref{tayl2}, we get the approximation
\begin{align}
\langle {v}^{\prime},{\Lambda}^{\prime} |&\hat{\Omega}\hat{\Theta}+\hat{\Theta}\hat{\Omega}|g(v,\Lambda)\rangle    \nonumber \\
& \simeq - [d_{{v}^{\prime}}(-4)y_{--}({v}^{\prime})\partial_{{\Lambda}^{\prime}} g({v}^{\prime}-4,{\Lambda}^{\prime})+d_{{v}^{\prime}}(4)y_{++}({v}^{\prime})\partial_{{\Lambda}^{\prime}} g({v}^{\prime}+4,{\Lambda}^{\prime})].
\end{align}
Since we are assuming that the only contributing values of $v$ are much bigger than 10, we can further introduce the approximations 
\begin{align}
\frac1{8}d_v(\pm4)y_{\pm\pm}(v)\simeq \pm\frac{1+\text{sign}(v\pm4)}{2}\sqrt{\frac{v}{2}\cdot\frac{v\pm 4}{2}}= \pm \tilde{y}_\pm(v),
\end{align}
from which we arrive at $\langle {v}^{\prime},{\Lambda}^{\prime}|\hat{\Omega}\hat{\Theta}+\hat{\Theta}\hat{\Omega}|g(v,\Lambda)\rangle  \simeq  8i \langle {v}^{\prime},{\Lambda}^{\prime}|  \hat{\tilde\Omega} |\partial_\Lambda g(v,\Lambda)\rangle $, where $|\partial_\Lambda g(v,\Lambda)\rangle$ is the state with wave function given by $\partial_\Lambda g(v,\Lambda)$.
In deducing this approximate identity, we have employed arguments similar to those explained in the previous paragraph and used that the action of the operator $\hat{\tilde\Omega}$, defined in \eqref{omegatilde}, is given by
\begin{align}
\hat{\tilde\Omega}|v\rangle= i \left[\tilde{y}_-(v)|v-4\rangle - \tilde{y}_+(v)|v+4\rangle\right].
\end{align}

Finally, and exactly as it was justified in \cite{hybrid-approx},  we approximate the action of the derivative $-4i\partial_\Lambda$ on functions of $\Lambda$ by the discrete derivative $\hat{\Theta}'$ [see \eqref{zetaprim}] at the scale $q_{\varepsilon}$  given in \eqref{scale2}.
As a result, we get $\langle {v}^{\prime},{\Lambda}^{\prime}|\hat{\Omega}\hat{\Theta}+\hat{\Theta}\hat{\Omega}|g(v,\Lambda)\rangle  \simeq - \langle {v}^{\prime},{\Lambda}^{\prime}|2 \hat{\tilde\Omega}\hat{\Theta}'| g(v,\Lambda)\rangle$,
as we wanted to prove. In summary, when the states $| g(v,\Lambda)\rangle$ are such that the dependence on $\Lambda$ of $g(v,\Lambda)$ is sufficiently smooth, and small values of $v$ (in terms of a fixed scale $v_m$) do not contribute, we can approximate the negative of \eqref{action-ani} by
\begin{align}\label{form315}
\langle {v}^{\prime},{\Lambda}^{\prime}|2 \hat{\tilde\Omega}\hat{\Theta}'| g(v,\Lambda)\rangle=&\frac{4}{q_\varepsilon}\big\{\tilde{y}_+({v}^{\prime})\left[g({v}^{\prime}+4,{\Lambda}^{\prime}+q_\varepsilon)-g({v}^{\prime}+4,{\Lambda}^{\prime}-q_\varepsilon)
\right]\nonumber\\-&    \tilde{y}_-({v}^{\prime})\left[g({v}^{\prime}-4,{\Lambda}^{\prime}+q_\varepsilon)-g({v}^{\prime}-4,{\Lambda}^{\prime}-q_\varepsilon)\right]\big\}.
\end{align}

If we consider states with $g(v,\Lambda)$ of the form \eqref{ani-v}, we expect that the corresponding anisotropy terms \eqref{form315} can be neglected in the Hamiltonian constraint, since the associated Gaussian profiles for the anisotropies are peaked on trajectories with vanishing momenta $\hat{\Theta}'$. Indeed, we get
\begin{align}
g(v\pm 4,\Lambda+q_\varepsilon)-g(v\pm 4,\Lambda-q_\varepsilon)=-g(v\pm 4,\Lambda)
2 \sinh\left(\frac{\sigma_s^2}{q_\varepsilon}[\Lambda-\bar\Lambda(v\pm 4)]\right) e^{-\frac{\sigma_s^2}{2} }.
\end{align}
The function $g(v\pm 4,\Lambda)$ only contributes when ${\sigma_s}[\Lambda-\bar\Lambda(v\pm 4)]
/{q_\varepsilon}$ is  $\mathcal{O}(1)$ (or smaller). Then, since $\sigma_s\ll \pi/2$, we obtain
\begin{align}
g(v\pm 4,\Lambda+q_\varepsilon)-g(v\pm 4,\Lambda-q_\varepsilon)\simeq -2g(v\pm 4,\Lambda) \sigma_s \times \mathcal{O}(1).
\end{align}
Therefore
\begin{align}
\langle {v}^{\prime},{\Lambda}^{\prime}|2 \hat{\tilde\Omega}\hat{\Theta}'| g(v,\Lambda)\rangle \simeq \frac{8\sigma_s}{q_\varepsilon}\big[\tilde{y}_-({v}^{\prime})g({v}^{\prime}-4,{\Lambda}^{\prime})-\tilde{y}_+({v}^{\prime})g({v}^{\prime}+4,{\Lambda}^{\prime})\big] \times \mathcal{O}(1).
\end{align}
Comparing now with the action of the FRW operator $\hat{\Omega}^2$,
\begin{align}
\langle {v}^{\prime},{\Lambda}^{\prime}|\hat{\Omega}^2| g(v,\Lambda)\rangle=-{y}_{--}({v}^{\prime})g({v}^{\prime}-4,{\Lambda}^{\prime})+2({v}^{\prime})^2g({v}^{\prime},\tilde{{\Lambda}^{\prime}})-{y}_{++}({v}^{\prime})g({v}^{\prime}+4,{\Lambda}^{\prime}),
\end{align}
we get that $ |\langle {v}^{\prime},{\Lambda}^{\prime}|2 \hat{\tilde\Omega}\hat{\Theta}'| g(v,\Lambda)\rangle| \ll  |\langle {v}^{\prime},{\Lambda}^{\prime}|\hat{\Omega}^2| g(v,\Lambda)\rangle|$ when  $\sigma_s\ll \pi/2$ and one stays in the sector $v>v_m$ under consideration (actually, as a consistency check, it is not difficult to see that both the expectation value and the dispersion of the anisotropy operator on the discussed states are negligible in comparison with those of the FRW operator).

This concludes the proof that the action of the anisotropy operator on states with profile \eqref{ani-v} can be disregarded, provided that $\sigma_s\ll \pi/2$ and that $N(v)$ is highly suppressed for $v\leq v_m$.

Let us now focus on the interaction term. In order to disregard this term,  the action of $\widehat{e^{-2\Lambda}}\hat\Omega^2$ on the considered states must be negligible. For the action of this operator we get
\begin{align}
\langle {v}^{\prime},{\Lambda}^{\prime}|\widehat{e^{-2\Lambda}}\hat{\Omega}^2| g(v,\Lambda)\rangle=&-{y}_{--}({v}^{\prime})e^{-2{\Lambda}^{\prime}}g({v}^{\prime}-4,{\Lambda}^{\prime})+2({v}^{\prime})^2e^{-2{\Lambda}^{\prime}}g({v}^{\prime},{\Lambda}^{\prime})
\nonumber\\&-{y}_{++}({v}^{\prime})e^{-2{\Lambda}^{\prime}}g({v}^{\prime}+4,{\Lambda}^{\prime}).
\end{align}
The first term on the right hand side only contributes when ${\sigma_s}[\Lambda-\bar\Lambda(v-4)]/ {q_\varepsilon}$ is  $\mathcal{O}(1)$, and the same can be said of the second and third terms when ${\sigma_s}[\Lambda-\bar\Lambda(v)]/ {q_\varepsilon}$ and ${\sigma_s}[\Lambda-\bar\Lambda(v+4)]/ {q_\varepsilon}$, respectively, are $\mathcal{O}(1)$. In addition, we have
\begin{align}
\label{3.19}
e^{-2\Lambda}g(v,\Lambda)=N(v)\exp\Bigg\lbrace{-\frac{\sigma_{s}^{2}}{2q_{\varepsilon}^{2}}\left[\Lambda-\bar{\Lambda}(v)+\frac{2q_{\varepsilon}^{2}}{\sigma_{s}^{2}}\right]^{2}}
\Bigg\rbrace\exp\left[-2\bar{\Lambda}(v)
+\frac{2q_{\varepsilon}^{2}}{\sigma_{s}^{2}}\right].
\end{align}
Note that this expression is valid also when evaluated at $v\pm4$.
	
Hence, if we assume that the dependence on $v$ of the function $\bar\Lambda(v)$ is smooth enough as to satisfy $\bar\Lambda(v)\simeq\bar\Lambda(v\pm 4)$ in the relevant region ($v>v_{m}$), then choosing 
$\bar\Lambda(v)\gg 1 \gg q_\varepsilon^2/\sigma_s^2$ for all values of $v>v_m$ guarantees that $| \langle {v}^{\prime},{\Lambda}^{\prime}|\widehat{e^{-2\Lambda}}\hat{\Omega}^2| g(v,\Lambda)\rangle| \ll  |\langle {v}^{\prime},{\Lambda}^{\prime}|\hat{\Omega}^2| g(v,\Lambda)\rangle|$.
	
Complementary, one can readily check explicitly that both the expectation value and the dispersion of $\widehat{e^{-2\Lambda}}\hat\Omega^2$ are negligible in our states under the above conditions. The proof follows exactly as in \cite{hybrid-approx}, using $\bar\Lambda(v)\simeq\bar\Lambda(v\pm 4)$.
	
In summary, for a quantum state such that its homogeneous gravitational part is given by $| g(v,\Lambda)\rangle=\sum_{v\in\mathcal{L}^+_{\varepsilon}}\sum_{\Lambda\in\mathcal{L}^{q_\varepsilon}_{\Lambda^\star}}g(v,\Lambda)|v,\Lambda\rangle$, with $g(v,\Lambda)$ defined in \eqref{ani-v}, the action of the operator $\hat{C}_\text{app}$ approximates that of the full Gowdy Hamiltonian constraint operator $\hat{C}_\text{G}$ provided that the following conditions are satisfied: 
\begin{itemize}
\item[i)] $N(v)$ has to be highly suppressed for $v\leq v_m$, with $v_m\gg 10$,
\item[ii)] $\sigma_s\ll \pi/2$,
\item[iii)] $\bar\Lambda(v\pm4)\simeq\bar\Lambda(v)\gg 1 \gg q_\varepsilon^2/\sigma_s^2$.
\end{itemize}
	
\subsection{Construction of the solutions}
\label{secB}

Let us now proceed to solve the constraint \eqref{constrainteq} for states \eqref{gen-states} with wave function
\begin{align}
\Psi(\phi,v,\Lambda,\mathfrak{n}^{\xi},\mathfrak{n}^{\varphi})&= \int_{-\infty}^{\infty} dp_{\phi}\, \psi(p_\phi)\chi(\mathfrak{n}^{\xi},\mathfrak{n}^{\varphi}) g(v,\Lambda) e_{p_\phi}(\phi).
\end{align}
Concerning the profiles $\psi(p_\phi)$ and $\chi(\mathfrak{n}^{\xi},\mathfrak{n}^{\varphi})$, we choose them as stated in Sec. \ref{iib}.
On these states, the only non-diagonal operator is $\hat{\Omega}^2$. For each value of the momentum of the homogeneous scalar field and of the modes occupancy numbers, the constraint equation reduces to ($g$ will also depend on the values of $p_\phi$ and $H_0$, but we obviate this dependence in our notation)
\begin{align}
(g(v,\Lambda) |\hat{\Omega}^2|{v}^{\prime},{\Lambda}^{\prime}\rangle=\left[\frac{4p_\phi^2}{3\pi G\hbar^2}+\frac{16}{3\beta}e^{2{\Lambda}^{\prime}} H_0(\mathfrak{n}^{\xi},\mathfrak{n}^{\varphi})\right]g({v}^{\prime},{\Lambda}^{\prime}),
\end{align}
which, taking into account the definition of $g(v,\Lambda)$ in \eqref{ani-v} and the properties required on the peaks of the Gaussian profiles for the anisotropies, leads within our approximations to the difference equation:
\begin{align}\label{diff}
N(v+4)&=\frac{1}{y_{++}(v)} \left[ 2v^2 - \frac{4p_\phi^2}{3\pi G\hbar^2}-\frac{16}{3\beta}e^{2\bar{\Lambda}(v)} H_0(\mathfrak{n}^{\xi},\mathfrak{n}^{\varphi})\right]N(v)\nonumber\\
&-\frac{y_{--}(v)}{y_{++}(v)} N(v-4).
\end{align}
Note that when $v=\varepsilon$ the last term vanishes, and therefore the solutions are {\emph{completely determined}} by the initial data $N(\varepsilon)$.

Among these solutions we are interested on those verifying conditions i), ii), and iii) of the previous section, as otherwise they would not provide approximate solutions to the full Gowdy constraint. We can impose conditions i) and iii) by choosing the function $\bar\Lambda(v)$ as follows
\begin{align}\label{lamb}
\bar\Lambda(v) =
\begin{cases}
\bar\Lambda_0, & \text{if }v\leq v_0, \\
h(v), & \text{if }v>v_0,
\end{cases}
\end{align}
for certain $v_0\geq v_m$, and such that $h(v\pm4)\simeq h(v)\gg 1 \gg q_\varepsilon^2/\sigma_s^2$ for all volumes $v>v_0$, and $\bar\Lambda_0=h(v_0)\gg 1 \gg q_\varepsilon^2/\sigma_s^2$. 

In this way, for $v\leq v_0$,  we are back to the case analyzed in \cite{hybrid-approx} with $\bar\Lambda_0$ constant and $f(v,\Lambda)=f(\Lambda)$, for which solutions are given by $N(v)=e^{\varepsilon}_{\rho(p_\phi,\bar\Lambda_0,\mathfrak{n}^{\xi},\mathfrak{n}^{\varphi})}(v)$. As we discussed in Sec. \ref{iib}, they are highly suppressed when $v\leq v_m= p_\phi/\sqrt{3\pi G\hbar^2}$, and we can always choose $p_\phi$ to be big enough as for $v_m$ to be much larger than 10, as we desire. Then, for $v>v_0$, relation \eqref{diff} gives deterministically the rest of the solution once it is supplemented with the input data $N(v_0-4)=e^{\varepsilon}_{\rho(p_\phi,\bar\Lambda_0,\mathfrak{n}^{\xi},\mathfrak{n}^{\varphi})}(v_0-4)$ and $N(v_0)=e^{\varepsilon}_{\rho(p_\phi,\bar\Lambda_0,\mathfrak{n}^{\xi},\mathfrak{n}^{\varphi})}(v_0)$. Also, note that throughout this period of the evolution (where $v>v_{0}$) these solutions remain peaked at $h(v)$. At dominant order with our approximations, the constraint equation just fixes the value of $N(v)$ via \eqref{diff} ($\forall v> v_0$) so that the considered states are solutions.

Finally, let us note that the scale $v_m$ used in the construction of the solutions has been defined above in an intrinsic way, in terms of a conserved quantity of the system, namely the momentum of the homogeneous scalar field, avoiding in this manner the introduction of an arbitrary parameter. 

\subsection{Different perfect fluid behaviors}

In view of the conditions that the approximations impose on the function $\bar\Lambda(v)$, or equivalently on  $h(v)$, let us now discuss whether we can choose that function in such a way that the solutions effectively behave as those of an isotropic flat FRW model coupled to a perfect fluid with equation of state $p=w \epsilon$, where $p$ and $\epsilon$ denote respectively the pressure and the energy density of the fluid, and where $w$ is the constant of proportionality between them. The constraint operator for such model reads
\begin{align}\label{const-appf}
\hat{C}_{\text{FRW+PF}}=-\frac{3\pi G\hbar^2}{8}\hat{\Omega}^2+\frac{\hat{p}_\phi^2}{2}+\alpha(1-w) \hat{v}^{1-w},
\end{align}
where PF stands for ``perfect fluid'' and $\alpha$ is a constant related to $\beta$. To deduce this equation we have employed that the matter contribution of a scalar field $\phi$ with potential $V(\phi)$ to the above constraint is
\begin{align}
\hat{C}_\phi=\frac{\hat{p}_\phi^2}{2}+V(\phi)a^6,
\end{align}
where the scale factor $a$ is proportional to $v^{1/3}$, and the fact that, for a perfect fluid, one has $V(\phi)=(\epsilon-p)/2$ and $\epsilon\propto a^{-3(1+w)}$ (see e.g. \cite{coluc}).

On the other hand, the states constructed in the previous section correspond to solutions of the approximate constraint 
\begin{align}\label{const-app2}
\hat{C}^\prime_{\text{app}}=-\frac{3\pi G\hbar^2}{8}\hat{\Omega}^2+\frac{\hat{p}_\phi^2}{2}+\frac{2\pi G \hbar^{2}}{\beta}e^{2\bar\Lambda(v)}\hat{H}_0,
\end{align}
where we are taking into account that, with respect to the anisotropy variable, those states are Gaussian and peaked on $\bar\Lambda(v)$.

Comparing both operators, we conclude that the analyzed states mimic a perfect fluid behavior if we choose 
\begin{align} 
\bar\Lambda(v) =
\begin{cases}
\ln \left[v_0^{(1-w)/2}\right], & \text{if }v\leq v_0 \\
\ln \left[v^{(1-w)/2}\right], & \text{if }v>v_0
\end{cases}
\end{align}
up to an additive constant, with $v_0\gg \exp \{2/(1-w)\}$. Note that, provided that we restrict our discussion to $w<1$, this function satisfies the requirements under equation \eqref{lamb}, needed for the validity of the approximations done in the previous section. In addition, we can also deal with the case $w=1$, which corresponds to the massless scalar field, already covered by the analysis of \cite{hybrid-approx}, as discussed at the end of Sec. \ref{sec:Gowdy}. Essentially, this case is reached by choosing the function $\bar{\Lambda}(v)$ equal to a constant and regarding the product $\alpha(1-w)$ in \eqref{const-appf} as a parameter that varies with $w$ and has a well defined limit when $w$ tends to the unity, which does not necessarily vanish.

This concludes the proof that anisotropic and inhomogeneous solutions of the Gowdy model can effectively behave as solutions of the Hamiltonian constraint of a flat FRW model coupled to a perfect fluid. Particularly interesting cases are dust, radiation, and a cosmological constant, since they may describe the dynamical behavior of our universe at different stages of its evolution (see e.g. \cite{coluc}). These cases are obtained simply by considering $w=0$, $w= 1/3$, and $w=-1$, respectively, in our formulas. Let us emphasize that such an effective description with a coupling to one of those perfect fluids with $w<1$ begins only when one reaches the volume $v_{0}$, while we find $w=1$ for smaller values of $v$. The phase with $w<1$ holds then indefinitely for $v>v_{0}$ by the very construction of the states that we have considered, as we explained in the previous section.

\section{Discussion}
\label{sec:conclu}

In this paper we have considered the hybrid quantization of the Gowdy $T^3$ model with linear polarization, LRS, and a minimally coupled massless scalar field \cite{hybrid-matter}. We have constructed approximate solutions (i.e., physical states) of this inhomogeneous model that in turn are solutions to the Hamiltonian constraint of a homogeneous and isotropic flat FRW model. The present analysis extends that of \cite{hybrid-approx}, which already provided approximate solutions of the Gowdy cosmology which behave dynamically (as far as the constraint is concerned) as those of the flat FRW space-times with a massless scalar field. Now, based on the approximations developed in \cite{hybrid-approx} and generalizing those previous results, we have constructed approximate solutions that behave as those of flat FRW coupled to a perfect fluid (with constant parameter $w$ at least from a certain instant of the evolution, namely,  from the instant when the volume $v$ reaches the value $v_0$ onwards). Our analysis is general enough to account for any perfect fluid with equation of state characterized by $w<1$ (as well as the already studied case $w=1$, which corresponds to the massless scalar field).

This analysis reflects the fact that specific quantum solutions of inhomogeneous models, in this case the Gowdy $T^3$ model, can resemble solutions of a flat homogeneous isotropic model with a particular isotropic matter content --at least for certain properties, like in this case the dynamical behavior imposed by the Hamiltonian constraint. Interestingly, those solutions are intrinsically far from being homogeneous, namely their anisotropies and  inhomogeneities are not negligible, and they can be made manifest by measuring generic observables related with those anisotropies and inhomogeneities. Even so, those states still behave in such a way that they lead to effective terms in the Hamiltonian constraint which are proper of a homogeneous model. More specifically, the particular solutions that we have constructed display a negligible momentum of the anisotropy and a negligible coupling between the homogeneous sector and the self-interaction of the inhomogeneities. As a result, they are solutions to the constraint of the flat FRW model coupled to a perfect fluid. This phenomenon is made possible by the quantum geometry effects characteristic of LQC (in particular by the exponential suppression of the eigenstates of the FRW geometry at small volume) and by the global behavior of the anisotropies and inhomogeneities on the considered quantum states, as far as the constraint is concerned. All these quantum effects invalidate Einstein's dynamics, which would have led to a singularity in the backwards evolution in which the anisotropic contributions would have acquired a dominant role.

It is worth emphasizing, once more, that the described perfect fluid behavior, in terms of an approximate homogeneous and isotropic Hamiltonian constraint, is not valid for generic quantum states of the Gowdy model. Generic states, solutions to the  constraint of our inhomogeneous and anisotropic model, do not possess the properties necessary for our approximations to hold. Therefore, as it should be obvious from the intrinsic inhomogeneous character of the Gowdy model, homogeneous and isotropic approximate descriptions would not be valid for generic quantum states, not even restricting all considerations to the Hamiltonian constraint. Our results in this work, nonetheless, show that the set of states in which descriptions of this type are possible are not so limited as one might have thought in principle, based just on the previos results of \cite{hybrid-approx}. Moreover, one might expect that the set of states in which such descriptions are approximately valid could be further extended beyond the family discussed here. Actually, the approximations proven in \cite{hybrid-approx} do not depend critically on the particular type of profile that the wave function takes on the anisotropy sector: it is only necessary that it is smooth and highly peaked on vanishing anisotropy momentum and in the region of large anisotropy variable $\Lambda$. Moreover,  the peaks can show any possible functional dependence on the FRW geometry as far as they respect that the sector of small volumes is suppresed. Based on these arguments, one can expect that the set of approximate solutions discussed in this work may indeed be enlarged. Consequently, even if the approximate homogeneous and isotropic dynamical behavior is specific of very special states, one can argue that the set of those states is not so severely limited after all.

On the other hand, the need to circumscribe our approximations to the kind of solutions explored here, at the end of the day, simply reflects the fact that effective descriptions reached within certain sets of quantum states generally depend on the particular set considered, via the specific correlations that exist on these states between the expectation values and quantum moments (dispersion and higher moments) of the phase space variables of the system. The states studied in this work certainly have correlations between the different moments, not only because the anisotropy profiles are Gaussian and peaked on vanishing anisotropy momentum, but also e.g. because the peak of each of these Gaussian wave functions has a particular dependence on the homogeneous volume $v$. General considerations about effective descriptions, like those discussed in \cite{effe2} for loop quantum gravity, are then overcome in the considered states owing to these correlations, a fact that explains why behaviors like those analyzed here are possible (actually, situations of this type are contemplated in \cite{effe2} if the correlations between the different moments are considered).

Although we have given full details of the conditions for the validity of our approximations to the Hamiltonian constraint of the Gowdy model, summarized in the three requirements imposed at the end of Subsec. \ref{iiia}, it is worth clarifying the mathematical sense in which the discussed states can be considered approximate solutions. First of all, the restiction to the sector of states where the region $v\leq v_m$ is suppressed, with $v_m\gg 10$, together with the smoothness on the anisotropies, allows one to approximate the action of the operators involved in the anisotropy term and in the term with the self-interaction of the inhomogeneities by simpler ones, as we have commented at the beginning of Subsec. \ref{iib}. This result was proven and discussed in depth in \cite{hybrid-approx}, and has been confirmed here (in Subsec. \ref{iiia}). The choice of Gaussian profiles made for the anisotropies is what differs now from the analysis of that reference, with peak trajectories that may depend on the volume $v$ in our present work. For these profiles, the conditions of a very small $\sigma_s$ (which can be regarded as a Gaussian width for the anisotropy momentum) and of a very large value of the peak of the anisotropy are simply necessary to disregard the two mentioned (and already approximated) terms of the constraint: the anisotropy term and that containing the self-interaction of the inhomogeneities. The neglected terms in this latter approximation can be treated as perturbations to our homogeneous and isotropic constraint. From our comments above, these perturbations should be negligible in an appropriate asymptotic limit of simultaneously vanishing Gaussian width $\sigma_s$ and anisotropy peak exponential $e^{-2\bar{\Lambda}(v)}$ [or, rather, of the product of the self-interaction of the inhomogeneities with $e^{-2\bar{\Lambda}(v)+2q_{\epsilon}^2/\sigma_s^2}$, in accordance with \eqref{3.19}, while the term $e^{2\bar{\Lambda}(v)} \hat{H}_0$ is kept finite]. Thus, the approximate solutions that we have discussed can be thought of as the leading contribution in a perturbative expansion, in which higher-order corrections would arise from the perturbations of the constraint. These considerations imply that the statement that our states provide approximate solutions, rather than local or confined to a certain interval in the dynamical evolution in terms of the volume $v$, is global inasmuch as the states correspond to perturbative solutions to differential or difference equations (see e.g. \cite{benderor}).

Our conclusions, extrapolated to the standard cosmological model with inflation, tentatively suggest the possibility that its phenomenology might be an effective description arising from solutions of a much richer underlying quantum model, with more degrees of freedom that organize themselves so that such a particular effective behavior emerges. At least from a conceptual point of view, this is a perspective that seems especially appealing and deserves further exploration, since it might reveal mechanisms associated with those additional degrees of freedom that could help to explain fundamental questions of cosmology, such as the existence of a cosmological constant or the origin of fields that produce inflation in the early Universe (see also \cite{oriti} for some recent ideas that partially share motivations of this type, and \cite{ale} for another analysis in which simple homogeneous descriptions are obtained from the study of specific types of quantum states that are inhomogeneous and whose properties leave imprints in the subleading corrections to the constraint).

\section{Acknowledgments}

The authors are grateful to C. Barcel\'o, L.J. Garay and D. Mart\'{\i}n de Blas for discussions. This work was partially supported by the Spanish MICINN/MINECO Projects No. FIS2011- 30145-C03-02 and FIS2014-54800-C2-2-P. 
M. M-B acknowledges financial support from the Netherlands Organisation for Scientific Research (NWO) (Project No. 62001772).

\end{document}